\documentclass[aps,twocolumn,footinbib,floatfix,showpacs]{revtex4}

\usepackage{graphicx}

\newcommand\kbar{{\bar k}}

\begin{document}
\title{Continuous Quantum Measurement and the Quantum to Classical Transition
       \vbox to 0pt{\vss
                    \hbox to 0pt{\hskip-50pt\rm LA-UR-02-5289\hss}
                    \vskip 25pt}
}
\preprint{LA-UR-00-XXXX}
\author{Tanmoy Bhattacharya}
\email{tanmoy@lanl.gov}
\homepage{http://t8web.lanl.gov/t8/people/tanmoy/}
\author{Salman Habib}
\email{habib@lanl.gov}
\homepage{http://t8web.lanl.gov/t8/people/salman/}
\author{Kurt Jacobs}
\homepage{http://t8web.lanl.gov/t8/people/kaj/}
\affiliation{T-8, Theoretical Division, MS B285, Los Alamos
National Laboratory, Los Alamos, New Mexico 87545}

\begin{abstract}
While ultimately they are described by quantum mechanics, macroscopic
mechanical systems are nevertheless observed to follow the
trajectories predicted by classical mechanics. Hence, in the regime
defining macroscopic physics, the trajectories of the correct
classical motion must emerge from quantum mechanics, a process
referred to as the quantum to classical transition. Extending previous
work [Bhattacharya, Habib, and Jacobs, Phys.~Rev.~Lett.~{\bf 85}, 4852
(2000)], here we elucidate this transition in some detail, showing
that once the measurement processes which affect all macroscopic
systems are taken into account, quantum mechanics indeed predicts the
emergence of classical motion.  We derive inequalities that describe
the parameter regime in which classical motion is obtained, and
provide numerical examples.  We also demonstrate two further important
properties of the classical limit.  First, that multiple observers all
agree on the motion of an object, and second, that classical
statistical inference may be used to correctly track the classical
motion.
\end{abstract}

\maketitle


\section{Introduction}
Macroscopic mechanical systems are observed to obey classical
mechanics to within experimental error.  However, the atoms which
ultimately make up these systems certainly obey quantum
mechanics.  Therefore, the question of how the observed classical
mechanics emerges from the underlying quantum mechanics arises
immediately.  This emergence, referred to as the quantum to classical
transition, is particularly curious in the light of the fact that the
equations of motion for the trajectories of classical mechanics are
nonlinear, and can therefore exhibit chaos, whereas even a proper
quantification of chaos in quantum mechanics has been difficult to
obtain~\cite{noqc}.

Note that the task of explaining the quantum to classical transition
(the QCT) is essentially a practical question: it is a question of
explaining why real systems, such as nonlinear pendulums, baseballs,
and other systems which can be be built and observed in the laboratory
obey classical mechanics (at least to within any experimental error).
It is not a question of obtaining classical mechanics precisely as a
formal limit of quantum mechanics.  In fact, due to the absence of
chaos in closed quantum systems~\cite{noqc}, and the non-commutativity
of the twin limits $\hbar\rightarrow 0$ (the semi-classical limit) and
$t\rightarrow\infty$ (the long-time limit necessary to describe chaos)
efforts to extract classical chaos as a formal limit have been less
than successful.

If one describes macroscopic objects sufficiently realistically using
quantum mechanics, then it should be possible to predict the (often
chaotic) trajectories of classical dynamics.  In order to do this, it
is important to realize that {\em all} real classical systems are
subject to interaction with their environment.  This interaction does
at least two things.  First, it subjects the system to noise and
damping~\cite{CL,qnoise} (as a consequence all real classical systems
are subject to noise and damping --- even if small), and second, the
environment provides a means by which information about the system can
be extracted (effectively continuously if desired), providing a
measurement of the system~\cite{cqm1,cqm1a,cqm2,cqm3,cqm4}.  

Two levels of description have been used to discuss the QCT. The first
utilizes the decoherence resulting from tracing over the environment
to suppress quantum interference \cite{deco}: it is assumed that no
dynamical information about the individual system has been extracted
from its environment. In many circumstances this alone can lead to an
effectively classical evolution of a phase space distribution
function~\cite{HSZ1998}. As mentioned above, a more fine-grained
description is achieved when the environment is taken to be a meter
that is continuously monitored, leading to a `quantum trajectory
unraveling' of the system density operator conditioned on the
measurement record. If one averages over all possible measurement
results, the description reverts to that at the level of phase space
distributions. However, the fine-grained description which explicitly
incorporates monitoring of the environment is required to understand
the QCT at the level of extracting classical trajectories from the
quantum substrate.

An example of an environment that naturally provides a measurement is
that of the electromagnetic field which surrounds the system.
Monitoring this environment consists of focusing the light which is
reflected from the system, allowing the motion to be observed.  If the
environment is not being monitored, then the evolution is simply given
by averaging over all the possible motions of the system.  Classically
this means an average over any uncertainty in the initial conditions,
and over the noise realizations.
However, in the absence of explicit observation (monitoring the
environment and recording the evolution) it is impossible to obtain
classical trajectories: the system must be 
described by an ever broadening probability (or pseudo-probability) 
distribution in phase space.  This is an experimental truism, and 
therefore applies regardless of whether the system is being treated 
by a classical or quantum mechanical theory.

Since all classical systems are subject to environmental interactions,
and since measurement is necessary to deduce the trajectories of
classical motion,
it may be expected that such environmental interaction, and the associated measurement process, will need to be included in a treatment 
that is adequate enough to predict the emergence of classical motion from
quantum mechanics. Indeed, recent work by a number of authors has 
made it increasingly clear that this provides a natural explanation
for the emergence of classical motion, and, therefore, a resolution of
the problem of the emergence of classical
chaos~\cite{Spiller,Graham,Percival,BHJ1,Milburn}.
Fortunately, the quantum theory of
environments and continuous measurement is now sufficiently well
developed that their effects can be treated in a fairly
straightforward manner, the emergence of classical dynamics verified,
and the mechanism of the quantum to classical transition elucidated.

Detailed studies of the QCT are particularly timely because current
experiments in quantum and atomic optics and condensed matter physics
are beginning to probe this transition directly in both ensemble and
individual system cases~\cite{expts}. Our approach here is to present
a general formalism for understanding the transition: more focused
analyses appropriate to specific experimental situations can easily be
developed based on this general approach.

In the following we examine the QCT in some detail.  In
Section~\ref{macobj}, we examine how macroscopic systems may be
treated, including environmental interactions and measurement.  In
Section~\ref{ineq}, we derive inequalities that describe the regime
under which classical motion emerges.  In Section~\ref{classest}, we
show that, in addition, classical state-estimation will 
work in the classical limit.  
In Section~\ref{numeg}, we provide two specific numerical
examples showing that classical motion is indeed obtained in the
regime predicted in Section~\ref{ineq}.  We finish with some
concluding remarks in Section~\ref{conc}.

\section{Describing the motion of macroscopic objects}\label{macobj}
A macroscopic object is composed of a very large number of quantum
degrees of freedom.  For example, we can consider the motions of the
atoms which comprise a massive object, and these are all coupled
together by the inter-atomic forces.  The equations of classical
mechanics are supposed to describe the dynamics of macroscopic
quantities, such as the center of mass; classical motion is not
observed in the entire many particle phase space.  Hence, one should
consider a change of variables, so as to write a Hamiltonian in terms
of the center-of-mass coordinate, $X$ (with conjugate momentum $P$).
This coordinate is coupled to all the other coordinates $x_i$ (in a
solid we might refer to these as the internal phonon modes, for
example).  Under the assumption that none of these environmental modes
is strongly perturbed by the dynamics, it is sufficient to treat them
as harmonic oscillators, and to couple them to the center-of-mass
motion via the Hamiltonian~\cite{CL,qnoise}
\begin{equation}
  H = H_{\mbox{\scriptsize cm}}(X,P) + \sum_{i} \left[ \frac{1}{2m_i}
          p_i^2 + \frac{\kappa_i}{2}(x_i - X)^2 \right]\, ,
\end{equation}
where $p_i$ are the momenta of the internal modes, and $\kappa_i$
gives the strength of the coupling between the center of mass and the
internal degrees of freedom.  The center of mass now constitutes
effectively an open system interacting with an environment consisting
of a large number of harmonic oscillators with a correspondingly large
range of frequencies.  This is the starting point for a treatment of
quantum Brownian motion, as developed by Caldeira and
Leggett~\cite{CL}.  Under general conditions, the phonon environment
can be treated as a heat bath, and in the double limit of weak
coupling to this bath and high temperature, it is possible to write a
very simple master equation for the center-of-mass
motion~\cite{qnoise}:
\begin{equation}
  \dot{\rho} = -\frac{i}{\hbar}[H_{\mbox{\scriptsize cm}}(X,P),\rho] -
                k_{env} [X,[X,\rho]]\, ,
\label{nomeas}
\end{equation}
where $k_{env}$ is determined by the $\kappa_i$ and the temperature.
This provides a simple description of the effects of the internal
degrees of freedom upon the macroscopic motion of an object for which
frictional effects are negligible, and the heating due to the noise is
not significant over time scales of interest.  If one wanted to treat
damped classical systems, then one would relax the weak coupling
approximation so as to give a master equation that explicitly contains
damping.  However, for simplicity, we will restrict our attention here
to classical Hamiltonian systems.

Another important environment which we need to consider is the quantum
electromagnetic field.  This interacts with the object, and provides a
natural mechanism for measurement of the center-of-mass position $X$.
In general, macroscopic objects are bathed in light from all
directions, and the light that is reflected may be monitored by a
large number of observers.  Since we are considering a one-dimensional
system, and since we wish to use the simplest description which
captures the essential aspects of the measurement process, we restrict
ourselves to interaction with an electromagnetic field in one
dimension.  In particular, we consider a laser reflected from the
object such that the phase shift provides information about $X$.
Performing an analysis of such a measurement, one finds that the
evolution of the system, conditioned upon the measurement record, may
be written as a stochastic master equation in the It\^o
formalism~\cite{Itoref} as~\cite{afm,measx}
\begin{eqnarray}
d\rho &=& - \{ \frac{i}{\hbar} [H_{\mbox{\scriptsize cm}}(X,P),\rho] + k
               ([X^\dag X,\rho]_+ - 2 X \rho X^\dag) \} dt \nonumber\\ 
      && + \frac{\sqrt\kbar}2 \{ (X\rho + \rho X^\dag) - \rho \mathop{\rm Tr} 
	\rho (X + X^\dag) \}  
	dW \nonumber\\
      &=& - \{ \frac{i}{\hbar} [H_{\mbox{\scriptsize cm}}(X,P),\rho] + k
	       [X, [X, \rho]] \} dt \nonumber\\
      && + \frac{\sqrt\kbar}2 \{ [X,\rho]_+ - 2 \rho \mathop{\rm Tr} \rho X \}
        dW \,,
\label{sme}
\end{eqnarray}
where the observed measurement record is given by
\begin{equation}
 dy = \mathop{\rm Tr} \rho X dt + \frac1{\sqrt\kbar} dW\, .
\label{measrec}
\end{equation}
In these equations, \(dW\) is a white noise generating a Wiener
process, $k$ gives the strength of the interaction between the light
and the object and is proportional to the power of the laser, whereas
\(\kbar\) gives the rate at which information about the system is
obtained.  When no information is obtained ({\it i.e.}, \(\kbar =
0\)), or if the measurement record is averaged over, the stochastic
master equation (\ref{sme}) reduces to the ordinary master equation
(\ref{nomeas}).

The ratio \(\eta \equiv \kbar / 8 k\), called the efficiency of the
measurement~\cite{cqm4}, is a measure of the fraction of the reflected
light that is actually detected by the observer in making the
measurement.  As will be clear from our discussion in Sec.~\ref{ineq},
though both \(k\) and \(\kbar\) arise from the interaction of the
system with the measurement environment, they play very different
roles: whereas \(\hbar^2k\) represents a noise on the system that
leads to spreading out in phase space, \(\kbar\) provides information
about the system leading to localization around individual
trajectories.  The fact that Eq.~(\ref{sme}) leads to a (completely)
positive evolution for all initial conditions if and only if \(\eta
\le 1\)~\cite{Barchielli} is a particular case of the general
information-disturbance principles in quantum mechanics: any process
that leads to information about a system must produce at least a
minimal unavoidable disturbance.

If there exist multiple observers dividing the available reflected
light up among them, then each sees an evolution with a value of
$\eta_i < 1$ (and, for positivity, \(\sum_i \eta_i \le 1\)), with a
different noise realization for each observer~\cite{MOcomment}. This is
certainly the case in reality, where each observer usually captures
only a small fraction of the available light.  In the regime in which
classical motion is obtained (which we will refer to as the {\em
classical limit}), all observers must agree on the motion of the
system to within experimental error, and we consider this question at
the end of the next section, and in our numerical examples.

Since the form of the equation resulting from interaction with the
internal modes is the same as that which results from failing to
monitor the light which is being used to probe the system, we can take
this environment into account in the same way that we take multiple
observers into account, that is, by taking an appropriate value of
$\eta<1$.  (The measurement constant $k$ is then adjusted to include
the contribution from $k_{env}$).

The stochastic master equation~(\ref{sme}) constitutes our description
of the evolution of the center-of-mass of a macroscopic object.  In
the following we will show that this description, while very simple,
is sufficiently realistic to obtain the correct classical motion in
the classical limit.  It should also be noted that while we have
chosen to measure the position $X$, the analysis which follows
suggests that the extraction of the classical limit is not sensitive
to the precise observables which are measured; as long as the
measurement provides sufficient information about the location of the
system in phase space, the classical limit will be obtained.  For
example, a continuous measurement of momentum will suffice, as long as
the forces on the system are spatially dependent.  In fact, other
authors have provided numerical support for this view by showing that
quantum state diffusion (using a measurement interaction which
includes damping)~\cite{Spiller} or a simultaneous measurement of
position and momentum~\cite{Milburn} are sufficient to induce the QCT
in the same manner.

If the correct classical mechanics is to be obtained, two conditions need
to be satisfied.  First, it must be possible to observe the system so
that its center of mass (and all other degrees of freedom considered
classical) is known sufficiently accurately on the scale of the
potential and relevant dynamical timescales.  Second, these
observed values, which we might identify as noisy counterparts of the
means of the sufficiently well localized distribution, $x\equiv\langle
X\rangle$ and $p\equiv\langle P\rangle$, should evolve according to
the classical Hamiltonian $H_{\mbox{\scriptsize cm}}(x,p)$ which has
the same functional form as the quantum Hamiltonian
$H_{\mbox{\scriptsize cm}}(X,P)$, with deviations small compared to
the classical scales.  In other words, the existence of the quantum to
classical transition implies that in the classical limit we can
replace the quantum operators with effective classical dynamical
variables,
\begin{equation}
   H(X,P) \rightarrow H(x,p)\, .
\end{equation}

\section{Inequalities governing the classical limit}\label{ineq}
We now ask for the parameter regime in which the evolution reduces to
classical motion.  As explained in the previous section this means
that the quantum distribution remains sufficiently localized (such
that the system can be said to be executing a trajectory), and that
this trajectory, characterized by $x\equiv\langle X\rangle$ and
$p\equiv\langle P\rangle$, follows that of the classical motion,
generated by $H_{\mbox{\scriptsize cm}}(x,p)$.

We proceed by first writing down the equations of motion for the first
and second moments of $X$ and $P$.  From Eq.~(\ref{sme}) these become
\begin{eqnarray}
   dx & = & \frac{p}{m} dt + \sqrt{\kbar} V_x dW \,, \label{eqx} \\
   dp & = & \langle F(X)\rangle dt + \sqrt{\kbar} C_{xp} dW \,,
            \label{eqp} 
\end{eqnarray}
and
\begin{eqnarray}
   dV_x    & = & \left[ \frac{2}{m} C_{xp} - \kbar V_x^2 \right] dt +
                 \sqrt{\kbar} K_{xxx} dW\,, \label{c1} \\ 
   dV_p    & = & \left[ 2\hbar^2 k - \kbar C_{xp}^2 + 2\partial_x F
                 C_{xp} \right] dt \nonumber  \\
           &   & + \partial_x^2 F K_{xxp} dt + \sqrt{\kbar} K_{xxp}
                   dW\,, \label{c2} \\  
   dC_{xp} & = & \left[ \frac{1}{m} V_p - \kbar V_xC_{xp} +
                           \partial_x F V_x \right] dt \nonumber \\  
           &   & + \frac{1}{2}\partial_x^2 F K_{xxx} dt +
                   \sqrt{\kbar} K_{xpp} dW \label{c3}\,, 
\end{eqnarray}
where
\begin{eqnarray}
   V_x    & = & \langle X^2 \rangle - \langle X\rangle \,, \\
   V_p    & = & \langle P^2 \rangle - \langle P\rangle \,, \\
   C_{xp} & = & \frac{1}{2}\langle XP + PX \rangle - \langle X\rangle
                \langle P\rangle \,, 
\end{eqnarray}
are the second cumulants, and the $K$'s are the third cumulants defined by
\begin{eqnarray}
   K_{abc} &=& \langle :ABC: \rangle - \langle :AB: \rangle 
             \langle C \rangle\nonumber - 
             \langle A \rangle \langle :BC: \rangle \\
   &&\qquad\qquad {} 
	     - \langle :AC: \rangle \langle B \rangle
             + 2 \langle A \rangle \langle B \rangle \langle C \rangle \nonumber\,,
\end{eqnarray}
where $A,B,C$ can be $X$ or $P$, and
the colons denote Weyl ordering of the operator products.  In the
above equations we use the simplified notation $F \equiv F(x) =
F(\langle X\rangle)$ and expand $F$ in a Taylor series about $X=x$
truncated to second order.  Without this truncation higher derivatives
of $F$ would appear in the equations for $V_p$ and $C_{xp}$,
multiplied by higher powers of the widths or by higher cumulants.
Truncating the power series for $F$ in this way is a good
approximation so long as the distribution is sufficiently localized
about $x$ and $p$.  Examining Eq.~(\ref{eqx}), one sees that to
maintain classical motion for $x$ one needs $\langle F(X) \rangle
\approx F(x)$, which happens when the system is localized enough so
that $V_x \partial_x F(x) \ll 2F(x)$.  It is the task of the
measurement to maintain such localization, and numerical
studies~\cite{BHJ1} have shown that it can indeed do so.

At this point, it is perhaps instructive to look at the origin of
localization of the individual trajectories.  The density matrix
obtained by solving Eq.~(\ref{sme}) is conditioned on the measurement
record Eq.~(\ref{measrec}), or equivalently, by the noise realization
\(dW\).  Averaging over these realizations results in the density
matrix of the unobserved system which can also be obtained by solving
Eq.~(\ref{nomeas}).  The second cumulants \(\sigma^2_{xx}\),
\(\sigma^2_{xp}\), and \(\sigma^2_{px}\) of that distribution are
related to the corresponding cumulants for each trajectory by the
relations:
\begin{eqnarray}
   \sigma^2_{xx} &=& \langle V_x \rangle_W + \mathop{\rm var}\nolimits_W
                        (x,x)\,, \nonumber\\
   \sigma^2_{xp} &=& \langle C_{xp} \rangle_W + \mathop{\rm var}\nolimits_W
			(x,p)\,, \nonumber\\
   \sigma^2_{pp} &=& \langle V_p \rangle_W + \mathop{\rm var}\nolimits_W
			(p,p)\,,
\end{eqnarray}
where \(\langle \cdot \rangle_W\) and \(\mathop{\rm
var}\nolimits_W(\cdot,\cdot)\) represent respectively the means and
(co-\nobreak)variances of the quantities, when considered as
distributions over trajectories.  The Wiener process damps the first
term on the right hand side of each equation by a term proportional to
\(-\kbar\), at the same time compensating this with a growth of the
last term in each equation.  As discussed in Sec.~\ref{classest}, this
is precisely the way in which classical measurement also selects well
defined trajectories out of an ensemble spreading out in phase space.

At the level of these second cumulants, the exact form of the damping
is, thus, immaterial.  The fact, however, that we derived these from
the stochastic master equation~(\ref{sme}) gives us not only a
theoretical `unraveling' of the master equation, but also provides a
physical meaning to each trajectory.  Furthermore, it guarantees that
the underlying evolutions of density functions are completely
positive, and, therefore, not only does the covariance matrix stay
positive, but also these equations can be completed into a hierarchy
of cumulant equations that automatically satisfy the appropriate
reality conditions.  Furthermore, as discussed in Sec.~\ref{macobj},
such a measurement process leads to the unavoidable noise proportional
to \(\hbar^2 k\) apparent on the right hand side of Eq.~(\ref{c2}); in
contrast, in the classical discussion in Sec.~\ref{classest}, the
corresponding noise term is unrelated to the measurement process and
can even be set to zero.  In fact, the truncation to the second
cumulants implies that no truly quantum effects of dynamics come into
play~\cite{habib} in our approximation, and the quantum scale
\(\hbar\) appears in our equations purely from this
information-disturbance consideration~\cite{footnote1}.

We will make two self consistent approximations in order to examine in
what regime classical dynamics emerges.  The first is to truncate the
power series in $F$ to second order.  The second is to neglect third
and higher cumulants in the equations for the second cumulants.  An
examination of the equations of motion for the third cumulants
appearing in Eqs.~(\ref{c1}) to (\ref{c3}) shows that indeed these are
damped by the measurement, again with damping coefficients
proportional to $\kbar$.  The fact that the wavefunction stays close
to Gaussian is also borne out by numerical simulations~\cite{BHJ1}.

Setting the third cumulants to zero in the equations for the second
cumulants, we solve for the stable steady-state:
\begin{eqnarray}
    V_x^{\mbox{\scriptsize ss}} & = & \sqrt{\frac{2
                C_{xp}^{\mbox{\scriptsize ss}}}{m\kbar}}\, , \label{Vxss} \\ 
   V_p^{\mbox{\scriptsize ss}} & = & mV_x^{\mbox{\scriptsize ss}} (
                \kbar C_{xp}^{\mbox{\scriptsize ss}} - \partial_x F)
                \,, \\ 
    C_{xp}^{\mbox{\scriptsize ss}} & = & \frac{\partial_x F}{\kbar}
                + \mbox{sgn}(m) \sqrt{\left( \frac{\partial_x
                F}{\kbar} 
	           \right)^2  +
                \frac{\hbar^2}{4\eta}} \,. \label{Cxpss}
\end{eqnarray}
where $\mbox{sgn}(m)$ is the sign of $m$ which we shall henceforth
take to be positive~\cite{footnote2} and $\partial_x F$ is taken to be
evaluated at a typical point in phase space.  Now, there are three
conditions that must be satisfied in order for the classical limit to
be obtained.  First, localization such that $V_x \partial_x F(x) \ll
2F(x)$, as discussed above, must be maintained, second, the noise
introduced by the measurement should be negligible compared to the
classical motion, and third, that the measurement record should follow
the motion of the position with sufficient accuracy.

Before examining these conditions in turn, two points are in
order.  First, localization and low noise are not really independent
constraints: in fact to provide effective damping for the covariance
matrix, the noise has to increase with increasing width of the state.
Conversely, noise also effects a spread in phase space of any
uncertain state, especially near unstable points.  It is convenient,
however, to treat the direct effect of the finite width of the state
on the deterministic evolution in a nonlinear potential as a question
of localization, and the rest as a question of low noise on the
trajectories.

As a second point, it is important to emphasize that the deviations of
the quantum trajectories from the classical ones can be different in
different parts of the phase space.  Nevertheless, in most
experimental situations, `classical' quantities evaluated are of
similar orders of magnitude almost everywhere, and, so, we shall
ignore these differences and consider them evaluated at a `typical'
point around the trajectory in question.

\subsection{Localization}
We start by noting that for the deterministic part of the equations of
motion for the quantum mean values $x$ and $p$ to match the classical
equations of motion, we need
\begin{equation}
    \langle F(X) \rangle = F(x) + \frac{1}{2}V_x \partial^2_x F(x) +
    \ldots
\end{equation}
to very closely approximate $F(x)$.  That is, we need
\begin{equation}
    r \equiv \left| \frac{\partial_x^2 F V_x}{2F} \right| \ll 1 \,.
\label{smallr}
\end{equation}
Replacing $V_x$ with its typical steady state value
[Eqs.~(\ref{Vxss},\ref{Cxpss})], we see that $r$ is the positive
solution to
\begin{equation}
    \frac{2m\kbar^2 F^2}{(\partial_x^2 F)^2} r^2 - (\partial_x F) =
        \sqrt{(\partial_x F)^2 + \frac{\hbar^2 \kbar^2}{4 \eta}} \,.  
     \label{loc1}
\end{equation}
Since this equation implies that $r$ is a monotonically decreasing
function of $\kbar$, Eq.~(\ref{smallr}) can provide a lower limit for
$\kbar$.  We examine this possibility in the following discussion.

Due to the positivity of its right hand side,
Eq.~(\ref{loc1}) implies that at the unstable point $\partial_x F >
0$, we must have  
\begin{equation}
   2 m \kbar^2 F^2 r^2 > (\partial_x^2 F)^2 \partial_x F\, .
\end{equation}
This alone means that to have $r\ll1$, it is necessary that
\begin{equation}
   \kbar^2 \gg \frac{(\partial_x^2 F)^2 |\partial_x F|}{mF^2} \,.
      \label{ineq13}
\end{equation}

Squaring Eq.~(\ref{loc1}) one sees that $r$ is the algebraically
largest solution of
\begin{equation}
   \kbar^2 = \frac{(\partial_x^2 F)^2}{mF^2 r^4} \left(
               \frac{(\partial_x^2 F)^2 \hbar^2}{16\eta m F^2} +
               \partial_x F r^2 \right) \,.  
\end{equation}
For this solution to be small would generically require 
\begin{equation}
   \kbar \gg \frac{(\partial_x^2 F)^2 \hbar}{4\sqrt{\eta} m F^2}
       \,, \label{ineq14} 
\end{equation}
except in the typical case of small nonlinearity
\begin{equation}
  (\partial_x^2 F)^2 \ll \frac{16 \eta m F^2|\partial_x F|}{\hbar^2}
            \, , \label{locineq1}
\end{equation}
when the width stays small at the stable points independent of the
value of $k$.  

Since, when the nonlinearity is large enough to violate
Eq.~(\ref{locineq1}), Eq.~(\ref{ineq14}) is stronger than
Eq.~(\ref{ineq13}), we can summarize these results as follows: If the
nonlinearity, characterized by $\partial_x^2 F$, is sufficiently weak
to satisfy Eq.~(\ref{locineq1}), then, at the unstable points
($\partial_x F > 0$) one needs
\begin{equation}
   8\eta k \gg \sqrt{\frac{(\partial_x^2 F)^2 |\partial_x
             F|}{2mF^2}} \,.  \label{locineq2}
\end{equation}
In the case of strong nonlinearity, we need
\begin{equation}
   8\eta k \gg \frac{(\partial_x^2 F)^2
      \hbar}{4\sqrt{\eta} m F^2}  \,,  \label{locineq3}
\end{equation}
to hold at all points.

\subsection{Low Noise}
In this subsection we consider the noise component of these equations.
In the classical limit the effect of this noise must be negligible on
the scale of the deterministic dynamics.  To compare the random noise
with the deterministic dynamics, we need to average over an
appropriate time scale: during a time $T$ upon which the dynamics is
effectively linear, the noise $dW$ provides an {\em rms} contribution
of $\sqrt{T}$.  We will define `low noise' to mean that the noise
contribution on this time scale is small compared to the deterministic
contribution.  The time scales upon which the dynamics is linear for
$x$ and $p$ are those in which the terms in the respective equations
do not change appreciably, and we will use the deterministic terms to
obtain these time scales.  The deterministic motion for $x$ is driven
by $p/m$, so appreciable changes occur when the change in momentum,
$\Delta p$, is of the order of $p$.  The resulting time scale for this
change is $T_x \sim |p/F|$.  The deterministic motion for $p$ is
driven by $\langle F(X)\rangle \approx F$.  Changes in $F$ are due to
changes in $x$: in particular $\Delta F \approx \partial_x F\Delta x$.
Hence the time scale for changes in $F$ is $T_p \sim
(m|F|)/(|p\partial_x F|)$.  Demanding that the change in $x$ and $p$
from the noise is small compared to that due to the deterministic
motion in these time intervals gives the two inequalities
\begin{eqnarray}
  \sqrt{\kbar} V_x & \ll & \frac{p}{m} \sqrt{T_x} =
               \sqrt{\frac{|p^3|}{m|F|}} \label{ncon1} \,, \\
  \sqrt{\kbar} C_{xp} & \ll & F \sqrt{T_p} = \sqrt{\frac{m|F^3|}{|p
               \partial_x F|}}\,.  \label{ncon2} 
\end{eqnarray} 
We will now examine these two inequalities in turn.

Considering the first inequality, and replacing $V_x$ with its typical
steady state value, given above, we have
\begin{equation}
   C_{xp}^{\mbox{\scriptsize ss}} \ll \frac{E|p|}{|F|} \, ,
\end{equation}
where $E = p^2/(2m)$ is the typical energy of the system.  Noting that
$E|p|/|F|$ has units of action, to simplify the following analysis we
will now define a dimensionless action $s$ by $s\hbar \equiv
E|p|/(4|F|)$.  Conceptually $s$ may be identified with the typical
action of the system in units of $\hbar$.  Using Eq.~(\ref{c3}) to
write $C_{xp}^{\mbox{\scriptsize ss}}$ in terms of $\kbar$ the
inequality becomes
\begin{equation}
   \sqrt{\left( \frac{\partial_x F}{\kbar} \right) +
      \frac{\hbar^2}{4\eta}} \ll 4 \hbar s - \frac{\partial_x
      F}{\kbar} \,.  \label{ineq1}
\end{equation}
The positivity of the left hand side immediately gives us the condition
\begin{equation}
    \frac{\partial_x F}{\hbar \kbar} < 4s \,.  \label{ineq6} 
\end{equation}
Now squaring both sides of Eq.~(\ref{ineq1}), and rearranging, we
obtain 
\begin{equation}
    \frac{\hbar^2}{4\eta} \ll 16\hbar^2 s^2 - \frac{8\hbar s
              \partial_x F}{\kbar} \,.  \label{ineq2}
\end{equation}
Because of Eq.~(\ref{ineq6}), this condition reduces to
\begin{equation}
   s \gg \frac{1}{8\sqrt{\eta}} \,, \label{ineq3}
\end{equation}
except at the unstable points ($\partial_x F > 0$) where one requires,
in addition, 
\begin{equation}
   \frac{\hbar\kbar}{\partial_x F} \gg \frac{32\eta s}{64\eta s^2 -
               1} \approx \frac{1}{2s} \,, \label{ineq8}
\end{equation}
where the approximate equality is implied by the inequality in
Eq.~(\ref{ineq3}).  

We now consider the inequality given by Eq.~(\ref{ncon2}). Replacing
$C_{xp}$ with its typical steady state value, and performing some
rearrangements we obtain
\begin{equation}
   \xi \ll 4\eta \left( s' - \mbox{sgn}(\partial_x F)
          \sqrt{\frac{4s'}{\xi}} \right) \,, \label{ineq4}
\end{equation}
where for compactness we have written
\begin{eqnarray}
   \xi &\equiv& \frac{\hbar\kbar}{|\partial_x F|}\nonumber\\
   \hbar s' &\equiv& \frac{mF^2|F|}{(\partial_x F)^2 |p|}\,.
\end{eqnarray}
Here $s'$ is a dimensionless quantity, which we will once again take
to be an estimate of the typical action of the system in units of
$\hbar$.  For $\partial_x F > 0$, the condition
\begin{equation}
      \xi \ll 4\eta \left( s' - \sqrt{\frac{4s'}{\xi}} \right)\, ,
      \label{ineq5}
\end{equation}
is satisfied whenever
\begin{equation}
   \frac{16}{s'} \ll \xi \ll 2\eta s' \, ,
     \label{ineq9}
\end{equation}
whereas, for $\partial_x F < 0$ (the condition is not useful when
$\partial_x F = 0$), it is sufficient that
\begin{equation}
   \frac{\hbar\kbar}{|\partial_x F|} \ll 4 \eta s' \,.
      \label{ineq10}
\end{equation}
Collecting all the inequalities in this subsection, ({\it i.e.} 
Eqs.~(\ref{ineq6}), (\ref{ineq3}), (\ref{ineq8}),
(\ref{ineq9}) and (\ref{ineq10})), we find that they are all
implied by
\begin{equation}
	\frac{ 2|\partial_x F| }{ \eta \bar s } \ll \hbar k \ll 
   	    \frac{ |\partial_x F|\bar{s} } {4}
   \,,
  \label{lnineq}
\end{equation}
where $\bar{s}\equiv\mbox{min}(s,s')$.  

\subsection{Faithful Tracking}
In the previous subsections we have been considering the motion of the
centroid of the quantum wave packet, $(x,p)$.  This centroid
represents the observer's true best-estimate of the mean value of
position and momentum at the current time, given the measurement
record.  To obtain this best estimate the observer must know the
dynamics of the system, given by $H_{\mbox{\scriptsize cm}}$, and then
integrate the full stochastic master equation, where the correct $dW$
is obtained continuously from the measurement record.

In practice, it is often merely the measured value of position which
is taken as the estimated value.  Hence, we need to find conditions
under which this value tracks the true best-estimate with sufficient
accuracy.  Since the measurement record in our formulation contains
white noise, the simplest way to model a realistic macroscopic
measuring apparatus is to low-pass filter, or {\em band limit} the
measurement record to obtain the continuous estimate of the position
(this is equivalent to making the reasonable assumption that all real
measuring devices have a finite response time).  This is achieved by
averaging the measurement record over some finite time $\Delta t$.  To
obtain an accurate estimate, $\Delta t$ must be short compared to the
dynamical time scale of the system.

If we assume that the change in $x$ over time $\Delta t$ is
negligible, then the error in the estimate of $x$ resulting from
averaging the measurement record $y(t)$ over $\Delta t$ is
\begin{equation}
   \sigma_{T}(x) = (8\eta k \Delta t)^{-1/2} \,.
\end{equation}
Hence, if to accurately track a classical dynamical system we require
a spatial resolution of $\Delta x$, and a temporal resolution of
$\Delta t$, then we must have
\begin{equation} 
  8 \eta k \ge \frac{1}{\Delta t (\Delta x)^2} 
\,.
\end{equation}
We also note, however, that in the observation of classical systems,
classical estimation theory is, in fact, often used to obtain the
classical equivalent of the quantum best-estimates provided by the SME
(Eq.~(\ref{sme})).  Such a procedure is most often used in classical
feedback-control applications.  In the classical limit, therefore,
such classical estimation procedures must work effectively, and we
will verify this in the next section.

\subsection{Summary}
We have now derived a set of inequalities which, when satisfied, lead
to the emergence of classical mechanics.  Consider first the
inequalities which come from the localization condition.  In the
macroscopic regime, which applies to common mechanical devices one
would build in the laboratory, the right hand side of inequality
(\ref{locineq1}) is extremely large compared to the typical
nonlinearity.  Consequently this inequality is satisfied, and the
resulting condition for $k$ is given by (\ref{locineq2}).  Note that
$\hbar$ does not appear in this inequality.  In fact, this is actually
a classical inequality, similarly required for classical continuous
measurement on classical systems.  In that case, the observer's state
of knowledge of the system is given by a classical probability density
in phase space, and this evolves as the system evolves and as
information is continuously obtained.

If the system is sufficiently small, and the nonlinearity
sufficiently large on the quantum scale so that inequality
(\ref{locineq1}) is not satisfied, then the condition for $k$ is
replaced by inequality (\ref{locineq3}).  This does contain $\hbar$,
and is, therefore, a uniquely quantum condition.  It appears due to
the unavoidable quantum noise which affects the dynamics strongly if the
nonlinearity is large on the quantum scale.

The left inequality in Eq.~(\ref{lnineq}), again is a classical
condition (the $\hbar$ arises because we chose to measure the action
$\bar s$ in units of $\hbar$): it reflects the observation that if the
measurement does not localize the motion, the state estimate changes
from moment to moment essentially randomly, or in other words, the
noise is large.  The right hand inequality in Eq.~(\ref{lnineq}) is
the direct effect of the irreducible noise coming from the measurement
process and is thus a quantum effect.  Together, as the action
increases, these low noise conditions put ever decreasing constraints
on the required measurement strength.

The faithful tracking condition is once again purely classical, in
that it also applies to classical observation.  It is simply the
condition on the accuracy of the measurement so that the measurement
record itself, as opposed to the estimated state, accurately tracks
the motion of the system from which the localization condition is
derived.

It is worth noting that the above inequalities also determine the
regime in which multiple observers agree on the motion of an object,
which is clearly an important property of the classical limit.  As
discussed in Section~\ref{macobj}, multiple observers can be taken into
account by giving each observer, $i$, a value of $\eta = \eta_i$ such
that $\sum_i \eta_i \le 1$, and giving each a different noise
realization, $dW_i$.  Furthermore, it is clear from the derivation of
the stochastic master equation~\cite{Barch93,DDZ} that the state
conditioned by the measurements made by all of the observers is
narrower than and consistent (in probability) with the state estimate
of each observer; {\it ipso facto}, the estimates of the different
observers must agree within errors.  Since the conditions derived in
this section can be satisfied with $\eta < 1$ (even with $\eta \ll
1$), and since these imply localization and accurate tracking of the
measurement record, under these conditions all observes will agree
upon the motion of the system to errors small on the classical scale.

\section{Classical Estimation in the Classical Limit}\label{classest}
When a classical system is subject to noise and continuous
observation, a classical theory of continuous state-estimation may be
developed to describe the continuous acquisition of information
regarding the system~\cite{Maybeck}.  Consider an observed classical
system whose dynamics is given by
\begin{equation}
    \left( \begin{array}{c} dx \\ dp \end{array} \right) = \left(
    \begin{array}{c} p/m \\ F_{\mbox{\scriptsize c}}(x) \end{array}
    \right) dt + \left( \begin{array}{c} 0 \\ \sqrt{2 g_p} \; dW_p
    \end{array} \right) 
\end{equation}
with measurement record 
\begin{equation}
    dy_{\mbox{\scriptsize c}} = x dt + \frac{dV}{\sqrt{g_m}} \,,
\end{equation}
{\it i.e.}, we consider a system with purely additive momentum noise
being observed continuously and with random errors. Here, $dW_p$ and
$dV$ are Wiener noises with $dV$ possibly correlated with the $dW_p$,
and $g_p$ and $g_m$ are positive real numbers.  Then the evolution of
the state of knowledge of the observer, described by a probability
density $P(x,p,t)$ obtained by averaging over $dW_p$ and conditioning
by $dy_c$, is~\cite{Maybeck,DHJMT}
\begin{eqnarray}
   dP & = & \left[ -(p/m)\partial_x - (F_{\mbox{\scriptsize c}} - g_p
            \partial_p)\partial_p \right] P dt \nonumber \\ 
      &   & + \sqrt{g_m} (x - \langle x \rangle) P dW \,,
\end{eqnarray}
where $dW = \sqrt{g_m}(x - \langle x \rangle) dt + dV$, and turns out
to be a Wiener noise, uncorrelated with the conditional probability
$P$.  Note that we can then write the measurement record for the
classical measurement as 
\begin{equation}
   dy_{\mbox{\scriptsize c}} = \langle x \rangle dt +
                 \frac{dW}{\sqrt{g_m}} \,, 
\end{equation}
and we see that this can be viewed as directly analogous to the
quantum measurement record.  The equations of motion for the classical
best estimates $\langle x \rangle_{\mbox{\scriptsize c}}$ and $\langle
p \rangle_{\mbox{\scriptsize c}}$, and the second order moments are 
\begin{eqnarray}
   d \langle x\rangle_{\mbox{\scriptsize c}} & = & \frac{\langle
                 p\rangle_{\mbox{\scriptsize c}}}{m} dt + \sqrt{g_m}
                 V_x dW \,, \label{eqxc} \\ 
   d\langle p\rangle_{\mbox{\scriptsize c}} & = & \langle
                 F_{\mbox{\scriptsize c}}(X)\rangle dt + \sqrt{g_m}
                 C_{xp} dW \,, \label{eqpc} 
\end{eqnarray}
and
\begin{eqnarray}
   dV_x    & = & \left[ \frac{2}{m}C_{xp} - g_m V_x^2 \right] dt +
                 \sqrt{g_m} K_{xxx} dW\,, \label{cc1} \\ 
   dV_p    & = & \left[ 2 g_p - g_m C_{xp}^2 + 2\partial_x F C_{xp}
                 \right] dt \nonumber  \\
           &   & + \partial_x^2 F K_{xxp} dt + \sqrt{g_m} K_{xpp} dW
                  \,, \label{cc2} \\  
  dC_{xp}  & = & \left[ \frac{1}{m}V_p - g_m V_xC_{xp} + \partial_x F
                 V_x \right] dt \nonumber \\ 
           &   & + \frac{1}{2}\partial_x^2 F K_{xxx} dt + \sqrt{g_m}
                 K_{xpp} dW \,.  \label{cc3} 
\end{eqnarray}
Identifying $g_m = \kbar$ and $g_p = \hbar^2 k$, we see that these
equations are identical to the quantum equations governing the
continuously estimated state [Eqs.~(\ref{eqx})-(\ref{c3})], and the
only way that quantum mechanics enters is in enforcing $\hbar^2 g_m
\leq 8 g_p$. Even though this is the case, it should be noted that
when the potential is nonlinear, the equations of motion for the
third and higher cumulants are not the same in the quantum and
classical cases, so in general the evolutions of the classical and
quantum estimates differ.  In the classical limit, however, the
conditional probability, or the state in the quantum case, is Gaussian
to a very good approximation so that the third cumulants can be set to
zero, and as a result they no longer feed into the equations for the
second order cumulants.  Consequently, the evolution of the classical
best estimates and second cumulants are identical to the quantum
estimates for the same measurement record, and as a result classical
estimation may be used to track dynamical systems in the classical
limit.

\section{Numerical examples}\label{numeg}
In this section we provide numerical support for the arguments in the
previous section.  We present two examples, and show that under the
conditions derived in the previous sections, the quantum wave packet
remains localized, the evolution of the centroid follows the classical
motion with negligible noise, and both the measurement record
(suitably band limited) and the classical state-estimate accurately
track the motion of the system for each of a set of observers.

To derive the equation of motion for the wavefunction of the
continuously observed system, assuming $N$ observers, one can first 
write down the Stochastic Schr\"odinger equation for the unormalised 
wavefunction for a single observer, making $N$ measurements. If the 
interaction strength for measurement $i$ is $\eta_i k$, then this is
\begin{eqnarray}
   d|\psi\rangle & = & \left[ -\frac{1}{\hbar}(iH(t) + \hbar k X^2) dt
                        \right.  \nonumber \\  
                 &   & \left.  + \sum_{i=1}^N 4 \eta_i k dr_i \right]
                        \vert\psi\rangle \,, 
   \label{dpsiN}
\end{eqnarray}
where the record for each measurement is given by 
\begin{equation}
  dr_i = \langle X\rangle dt + \frac{dW_i}{\sqrt{8\eta_i k}} .
\end{equation}
Now we let each observer have access to just one of the measurement 
records. In addition, we choose $\sum_i\eta_i = 1$, so that $\eta_i$ 
represents the fraction of the total measurement interaction strength 
$k$ used by each observer.  The evolution 
of the state-of-knowledge for any particular observer (who only has 
access to her measurement record) can be calculated by averaging over 
the noise realizations for all the other observers while keeping the 
measurement record for the observer in question fixed. The resulting 
equation of motion for the state-of-knowledge of observer $i$, given 
the measurement record $dr_i$ generated by the stochastic Schr\"odinger 
equation (\ref{dpsiN}) is the stochastic master 
equation~\cite{Barch93,DDZ}
\begin{eqnarray}
  d\rho & = & -\frac{i}{\hbar}[H,\rho] dt - k [X,[X,\rho]] dt
        \nonumber \\ & & + ( [X,\rho]_+ - 2 \rho \mathop{\rm Tr}[\rho
        X]) \sqrt{2\eta_i k} dV_i \label{smemult}
\end{eqnarray}
where
\begin{equation}
  dV_i = \sqrt{8\eta_i k}(dr_i - \mathop{\rm Tr}[\rho X] dt) .
\end{equation}
Note that this is, in fact, just Eq.(\ref{sme}), because as far as 
observer $i$ is concerned, all the other observers are simply gathering 
part of the environment to which $i$ has no access. In addition, 
the fractions $\eta_i$ are the respective measurement efficiencies. 

To simulate multiple observations on a given system, we first
integrate the stochastic Schr\"odinger equation which generates a set
of measurement records, one for each observer.  We then integrate the
corresponding stochastic master equations using the measurement record
for each observer.  The state-of-knowledge of each observer over time
can then be compared to the `actual' evolution of the system state
vector given by Eq.~(\ref{dpsiN}). The stochastic Schr\"odinger and
master equations were integrated in time using a spectral
split-operator method.  Since the classical limit is obtained when the
extent of the wavefunction is small compared to the range of motion of
the centroid, the algorithm is designed so that the computational grid
follows the wavefunction in both position and momentum space, and this
is crucial for efficient computation.

In treating systems of different sizes and actions it is convenient to
choose units for the system variables to keep the numerical value of
the action close to unity.  Due to this system dependent choice of
units, the fixed quantity $\hbar$ has a system dependent numerical
value; and indeed we expect the classical limit when $\hbar \ll 1$ in
these units.  This is what we demonstrate below.

\begin{figure}
\leavevmode\includegraphics[width=0.9\hsize]{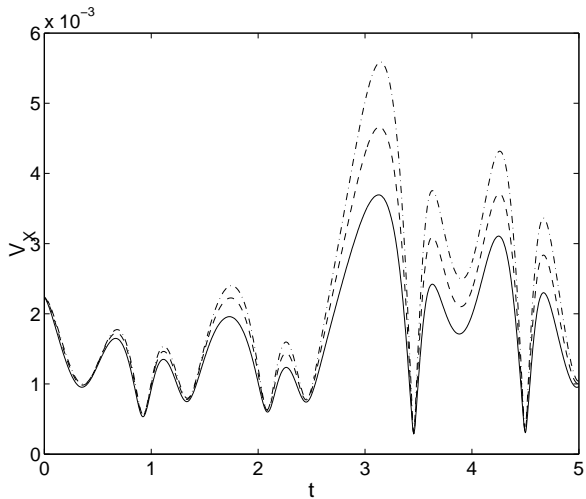}
\caption{The standard deviations of the state estimates for each of
the three observers for the Duffing oscillator, plotted over a
duration of $t=5$.  Solid line: observer with $\eta=0.5$; dashed line:
observer with $\eta=0.3$; dash-dot line: observer with $\eta=0.2$.}
\label{fig1}
\end{figure}

\subsection{The Duffing oscillator}
The Duffing oscillator is a sinusoidally driven double-well potential,
with Hamiltonian
\begin{equation}
   H(t) = \frac{P^2}{2m} + BX^4 - AX^2 + \Lambda X \cos(\omega t) \,.
\end{equation}
We choose $m=1$, $A=10$, $B=0.5$, $\Lambda=10$ and $\omega=6.07$.  At
times when the driving is zero, this puts the minima of the two
potential wells at $\sim\pm 3.2$, with a central barrier height of
$50$.  We choose $\hbar=10^{-5}$ and $k=10^5$, which is sufficient to
satisfy the inequalities derived in Section~\ref{ineq} for all but
tiny values of $\eta$, and therefore puts the system in the classical
regime.  We now evolve the system with three observers, and set their
measurement efficiencies to be $0.5,0.3,0.2$ respectively.  We first
calculate the position variance $V_X$ of the wave-function given by
evolving Eq.~(\ref{dpsiN})), and verify that this remains sufficiently
small.  Running the simulation for a duration of $t=5$, the maximum
value of $\sqrt{V_X}$ is $2.7\times 10^{-3}$, and the {\em rms} value
over the evolution is $1.4\times 10^{-3}$.  The localization condition
is therefore well satisfied, and an inspection of the evolution of the
centroid shows that the noise is indeed negligible.  Showing that the
evolution is indeed the classical evolution is more nontrivial since
the system is chaotic: any small difference in the noise on two
trajectories will cause them to diverge rapidly, and one cannot
therefore simply compare the trajectory to the equivalent noise-free
classical trajectory.  In Ref.~\cite{BHJ1}, the classical dynamics was
verified by comparing the stroboscopic map and the largest Lyapunov
exponent obtained from the quantum evolution and their classical
equivalents.  Here we calculate the continuously estimated state, both
quantum and classical, for the different observers, and show that
these agree, and agree between observers.

\begin{figure}
\hbox{%
\hfill
\includegraphics[width=1.0\hsize]{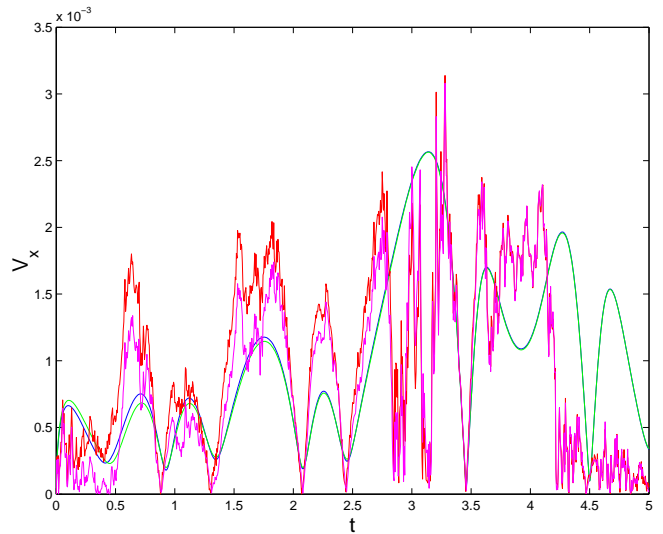}
}
\caption{Plot of the error standard deviation (blue) and the
difference between the estimated and true means for a single noise
realization (red) for the simulation of the Duffing oscillator using
the stochastic master equation with $\eta=0.5$. The green and purple
curves plot the same quantities for the Gaussian estimator.}
\label{fig2}
\end{figure}

We now calculate the quantum state-estimate for each observer,
obtained by integrating Eq.~(\ref{smemult}), and compare this with the
classical (Gaussian) estimate for each observer, obtained by
integrating Eqs.~(\ref{eqxc})-(\ref{cc3}).  In Figure~\ref{fig1} we
plot the uncertainty in position (characterized by $\sqrt{V_X} =
\sqrt{\mbox{Tr}[\rho X^2] - \mbox{Tr}[\rho X]^2}$) for the quantum
state estimated by each observer over the duration of the run.  All
these remain small.  The {\em rms} of $\sqrt{V_x}$ for each observer
over the duration of the run is given in Table~\ref{tab1}.  The
evolution of the uncertainty in position for the Gaussian
state-estimate is essentially identical to the quantum estimate, and
the {\em rms} of $\sqrt{V_X}$ for this estimate is also given in
Table~\ref{tab1}.  Note that the position variances for each observer
are, as expected, larger than the variance of the wave-function
calculated using the stochastic Schr\"odinger equation. Since the
solution to Eq.~(\ref{dpsiN}) can be viewed as an `unraveling' of the
stochastic master equation in Eq.~(\ref{smemult}), the difference
between the variances of the `true' state estimate from the former and
the individual observers' state estimate from the latter provides an
estimate of the amount by which their means differ, averaged over
noise realizations for all the observers. We will refer to this,
for want of a better name, as the error variance, and its square root
as the ``error standard deviation''.  The {\em rms} value of this
error standard deviation is $1.2\times 10^{-3}$, $1.7\times 10^{-3}$
and $2.1\times 10^{-3}$ for observers with $\eta=0.5$, $0.3$ and $0.2$
respectively for both the stochastic master equation and Gaussian
simulations.  In Figure~\ref{fig2} we plot the evolution of the error
standard deviation, and also the actual difference between the
estimated mean and the mean of the wave-function, for both the master
equation simulation and the Gaussian estimator. The small difference
between these last two is most probably due solely to the fact that
the mean of the computationally intensive master equation simulation
has not completely converged at the value of the time step
employed. (This difference is not seen for the computationally simpler
delta-kicked rotor system discussed in the next subsection.)

Each observer may also track the position simply by averaging her
measurement record over a suitable time period ({\it i.e.}, by low
pass filtering the measurement record).  Naturally this period should
be as long as possible so as to filter out the noise, but short enough
so as not to filter out the deterministic motion.  For this system we
use a time period of $2.5\times 10^{-2}$ for the filtering.  The
average {\em rms} deviation of this estimate from the mean position of
the wave-function is also given in Table~\ref{tab1} for each observer.
From this we see that all observers can effectively track the motion
of the particle (up to an error in position of about $10^{-2}$) using
their measurement records directly.

\begin{table}
\caption{{\em rms} standard deviation of state-estimates for the
Duffing oscillator and the {\em rms} deviation of the averaged
measurement record.}
\begin{tabular}{|l@{\hspace{0.2cm}}|l@{\hspace{0.2cm}}|l@{\hspace{0.2cm}}|l@{\hspace{0.2cm}}|}
   \hline
   {\bf Observer's $\eta$}& $\eta=0.5$ & $\eta= 0.3$ & $\eta = 0.2$ \\ 
   \hline 
   {\bf Quantum } & $1.9\times 10^{-3}$ & $2.3\times 10^{-3}$ &
                           $2.6 \times 10^{-3}$ \\
   {\bf Gaussian } & $1.9\times 10^{-3}$ & $2.3\times 10^{-3}$ & 
                           $2.6 \times 10^{-3}$ \\
   {\bf Averaged Record} & $8.2\times 10^{-3}$ & $9.3\times 10^{-3}$ &
                           $1.1\times 10^{-2}$ \\ 
   \hline
\end{tabular}
\label{tab1}
\end{table}

\subsection{The delta-kicked rotor}

\begin{figure}
\leavevmode\includegraphics[width=0.9\hsize]{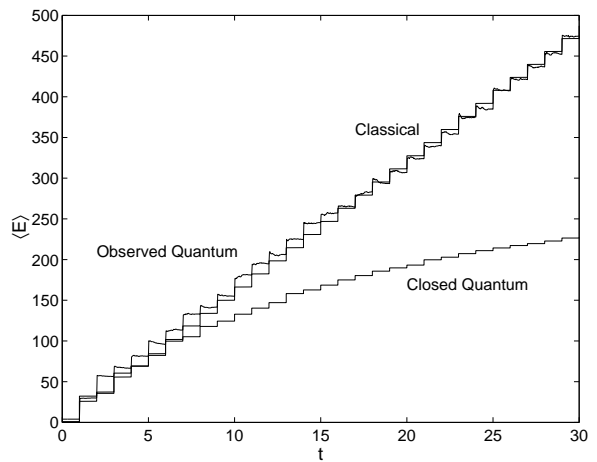}
\caption{The average kinetic energy for the delta-kicked rotor as a
function of time.  The classical value is obtained by averaging over
10,000 trajectories.  The observed quantum value was obtained by
averaging over 1000 trajectories.}
\label{fig3}
\end{figure}

The delta-kicked rotor obeys the Hamiltonian
\begin{equation}
   H(t) = \frac{P^2}{2m} + \kappa \cos(X) \sum_{n=0}^\infty\delta(t-n)
   \,.
\end{equation}
It is, thus, a free particle, which experiences regular kicks from the
potential of a nonlinear pendulum. For a wide range of parameters, the
quantum behavior of this system (by which we mean the evolution of the
closed system) is very different from the classical motion.  In
particular, after a few kicks the average energy of the closed
classical system increases linearly with time.  In the closed quantum
system, however, the average energy reaches a maximum value and after
that point remains fairly constant. This is termed dynamical
localization. We now simulate the evolution of the observed
wave-function for this system, with the same values of $\hbar$ and $k$
as we used for the Duffing oscillator, and with the same three
observers.  For the system parameters we will choose $\kappa=10$ and
$m=1$, and integrate for a time period of 30 kicks.  First we check
the localization of the wave-function given by integrating the
stochastic Schr\"odinger equation, and find that the average value of
$\sqrt{V_X}$ is $2.1\times 10^{-3}$, and the maximum value obtained
during the run is $3.2\times 10^{-3}$.  We check that the mean energy
is indeed behaving in a classical fashion by averaging this energy
over many realizations, and comparing this to the classical value.  In
Figure~\ref{fig3} we plot the average energy of the observed quantum
system, using $\hbar=0.1$ and $k=10$, along with both the
classical result and the quantum result for $\hbar=0.1$.

\begin{figure}
\hbox{%
\hfill
\includegraphics[width=1.0\hsize]{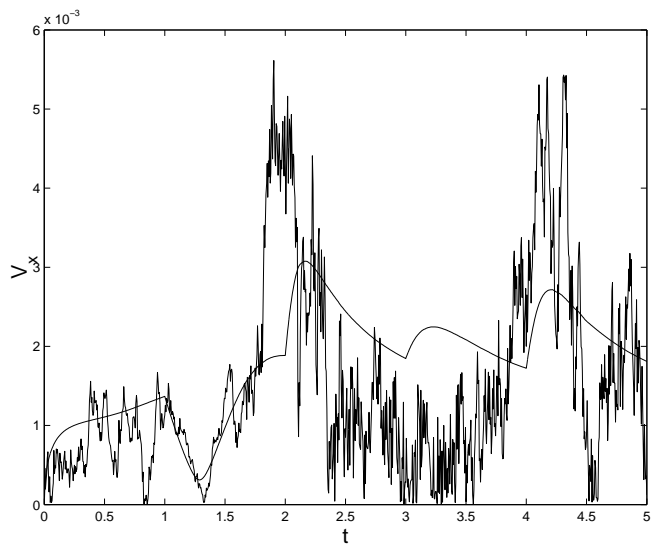}
\hfill
}
\caption{Plot of the error standard deviation (green) and the
difference between estimated and true means for a single noise
realization (red) for a simulation of the delta-kicked rotor using a
stochastic master equation with $\eta=0.5$. The results for the
Gaussian estimator are indistinguishable on this scale.}

\label{fig4}
\end{figure}

We next compare the position uncertainties in the state-estimates of
the different observers, as above for the Duffing oscillator, and
present these results in Table~\ref{tab2}. The Gaussian estimator
agrees with the stochastic master equation, and the uncertainties are
small, so that the observers effectively all agree on the motion.  The
averaged measurement record also tracks the motion effectively.  The
{\em rms} value of the error standard
deviation is $1.9\times 10^{-3}$, $2.9\times 10^{-3}$ and $3.7\times 10^{-3}$
for observers with $\eta=0.5$, $0.3$, and $0.2$ respectively. In
figure~\ref{fig4} we plot the evolution of the error standard
deviation for the observer with $\eta=0.5$ and also the actual
difference between the estimated mean and the mean of the
wave-function for a single realization of the stochastic master
equation simulation. The equivalent plots for the Gaussian estimator
are virtually indistinguishable.

To conclude, we see from the above simulations that (i) in the
classical regime the full quantum state-estimation reduces to Gaussian
state-estimation, and hence classical state-estimation may be used,
(ii) even without the use of true (and therefore optimal)
state-estimation, low pass filtering of the measurement record alone
provides adequate tracking of the system, and (iii) since the errors
in the respective estimates are small, all observers effectively agree
upon the motion of the system.

\begin{table}
\caption{Average deviation of state-estimates for the delta-kicked
rotor and the {\em rms} deviation of the averaged measurement record.}
\begin{tabular}{|l@{\hspace{0.2cm}}|l@{\hspace{0.2cm}}|l@{\hspace{0.2cm}}|l@{\hspace{0.2cm}}|}
    \hline 
    {\bf Observer's $\eta$}& $\eta=0.5$ & $\eta= 0.3$ & $\eta = 0.2$ \\
    \hline
    {\bf Quantum } & $2.9\times 10^{-3}$ & $3.6\times 10^{-3}$ &
             $4.3 \times 10^{-3}$ \\ 
    {\bf Classical } & $2.9\times 10^{-3}$ & $3.6\times 10^{-3}$ & 
             $4.3 \times 10^{-3}$ \\ 
    {\bf Averaged Record } & $8.6\times 10^{-3}$ & $1.0\times 10^{-2}$ &
             $1.1\times 10^{-2}$ \\ \hline
\end{tabular}
\label{tab2}
\end{table}

\section{Conclusion}\label{conc}

The emergence of classical dynamics remains a central issue in
understanding the predictions of quantum mechanics, especially now
that experiments are becoming available to probe this transition
directly~\cite{expts}.  In this paper, by deriving general
inequalities which determine when classical mechanics will emerge, and
by providing numerical examples, we have presented very substantial
evidence that quantum measurement theory provides a completely
satisfactory answer to the question of how classical mechanics, and
hence classical chaos, emerges in a quantum world.  In doing so we
have shown in some detail how the mechanism for this transition can be
understood as a result of localization and noise suppression in the
classical regime.

While the emergence of classical dynamics for a single motional degree 
of freedom now appears to be well understood, the quantum to classical 
transition as yet holds many unanswered questions. What happens, for 
example, to the dynamics of a system as it passes ``through'' the 
transition? How do systems behave when they are neither fully quantum 
nor fully classical? For example, it is known that the delta kicked rotor demonstrates a complex behavior in the transition region~\cite{BHJS}. 
Further questions include how classical dynamics emerges for other 
degrees of freedom, such as spin, and what happens, for example, when 
spin and motional degrees of freedom are coupled? Must all the subsystems 
have a large action (we note that this has recently been investigated, see~\cite{GADBHJ}), and must all the degrees of freedom be 
continuously measured, or will a subset suffice? For a spin system, 
must one measure all the components of spin, or will a single 
component suffice? Fortunately we are now at the point where one can 
not only pose these questions, but expect that solid answers will soon 
be forthcoming.


\begin{thebibliography}{99}
\bibitem{Barchielli} The `if' part of the statement is clear from
  early work in Ref.~\protect\cite{cqm1}.  The `only if' is clear by
  noting that if \(\rho\) is an one-dimensional projector (so that
  \(\mathop{\rm Tr} \rho^2 = 1\)) that is not into an eigenspace of
  \(X^\dag X\), \(\eta > 1\) implies \(d \mathop{\rm Tr} \rho^2/dt\)
  averaged over the noise \(d W\) is positive definite.
\bibitem{MOcomment} 
  This follows most simply by treating each observer as having 
  access to a separate output channel, or environment, but it is 
  clear from~\cite{HMW} that one could obtain the same result 
  by dividing up the available light using a sequence of beam 
  splitters. 
  For an explicit treatment of multiple observers see 
  reference~\cite{Barch93}. A recent and less mathematical treatment 
  is also given in reference~\cite{DDZ}.
\bibitem{footnote1} Positivity at this second cumulant level does not
  require the noise term: the covariance matrix stays positive even
  without the addition of this piece.
\end{thebibliography}
\end{document}